\documentclass[preprint,aps,tightenlines,floatfix,showpacs]{revtex4}

\usepackage{graphicx}
\usepackage{bm}
\usepackage{amsmath}
\usepackage{amssymb}

\begin{document}
\title{Collective-coupling analysis of spectra 
of mass-7 isobars: $^7$He, $^7$Li, $^7$Be, $^7$B.}

\author{L. Canton$^{(1)}$}
\email{luciano.canton@pd.infn.it}
\author{G. Pisent$^{(1)}$}
\email{pisent@pd.infn.it}
\author{K. Amos$^{(2)}$}
\email{amos@physics.unimelb.edu.au}
\author{S. Karataglidis$^{(2,3)}$}
\email{s.karataglidis@ru.ac.za}
\author{J. P. Svenne$^{(4)}$}
\email{svenne@physics.umanitoba.ca}
\author{D. van der Knijff$^{(5)}$}
\email{dirk@unimelb.edu.au}

\affiliation{$^{(1)}$Istituto Nazionale di Fisica Nucleare, sezione di Padova, \\
e Dipartimento di Fisica dell'Universit$\grave {\rm a}$
di Padova, via Marzolo 8, Padova I-35131,
Italia}
\affiliation{$^{(2)}$School of Physics, University of Melbourne,
Victoria 3010, Australia}
\affiliation{$^{(3)}$Department of Physics and Electronics, Rhodes University,
Grahamstown 6140, South Africa}
\affiliation{$^{(4)}$
Department of Physics and Astronomy,
University of Manitoba,
and Winnipeg Institute for Theoretical Physics,
Winnipeg, Manitoba, Canada R3T 2N2}
\affiliation{$^{(5)}$Advanced Research Computing, Information Division,
University of Melbourne, Victoria 3010, Australia}

\date{today}

\pacs{24.10-i;25.40.Dn;25.40.Ny;28.20.Cz}

\begin{abstract}
A nucleon-nucleus interaction model has been applied to ascertain the
underlying character of the negative-parity spectra of four isobars of
mass seven, from neutron-- to proton--emitter driplines.  
With one and the same nuclear potential defined by a simple coupled-channel model,
a multichannel algebraic
scattering approach (MCAS) has been used to determine the bound and resonant 
spectra of the four nuclides, of which $^7$He and $^7$B are particle
unstable.  Incorporation of Pauli blocking in the model
enables a description of
all known spin-parity states of the mass-7 isobars.
We have also obtained spectra of similar quality
 by using a large space no-core shell model.
Additionally, we have
studied  ${}^7$Li and  ${}^7$Be using a dicluster model.
We have found   a dicluster-model potential that can reproduce
the lowest four states of the two nuclei,
as well as the relevant
low-energy elastic scattering cross sections. But, with this model, 
the rest of the
energy spectra cannot be obtained.
\end{abstract}

\maketitle

\section{Introduction}
Currently there is much interest in the structure of, and reactions with,
radioactive nuclei. In particular, attention has been given to 
weakly-bound light nuclei, which may manifest exotic structure. 
The very extended neutron matter distributions of ${}^6$He and ${}^{11}$Li
that have been called neutron halos are examples. They  contrast with 
${}^8$He and ${}^9$Li which have neutron skins.  Signatures 
of those neutron matter distributions have been noted in the
cross sections from elastic and inelastic scattering  of the nuclei
from hydrogen~\cite{Ka00,St02}.
As yet, relatively few such experiments have been made (elastic and inelastic  
scattering of radioactive ion beams (RIB) from a hydrogen target). 
We hope that will change in the near future, since
appropriate scattering theories to describe the events not only exist but also
have been 
implemented~\cite{Am03,St02}. Most of the existing RIB-hydrogen scattering data have been 
taken with ions of medium energies, analyses of which are appropriately
made using  a $g$-folding model of the optical potential
 for elastic scattering and a distorted wave approximation (DWA) analysis of 
inelastic scattering data~\cite{Am00}. For low energies, however,
a coupled-channel theory is more relevant. Then  the MCAS method~\cite{Am03} is a most 
appropriate
way to proceed. 

Low energy nucleus-nucleon scattering data exhibit resonances,  analyses of
which reflect structure of the compound nucleus formed by the 
ion and proton target. Such was described in a recent publication~\cite{Ca06}.
Also MCAS analyses of data from the scattering of low energy 
${}^{14}$O ions from hydrogen~\cite{Gu05} revealed that 
the spectrum should have a number of narrow resonances at energies slightly higher 
than the observed two broad resonances  that
coincide with the ground and first excited states of the proton-unstable 
${}^{15}$F.
 The centroids, widths, and spin-parity assignments all are
determined using the MCAS approach~\cite{Ca06}.

 Herein we use the MCAS theory for the interactions of ${}^6$He and of ${}^6$Be 
with nucleons.  The spectrum of ${}^7$Li is used to set that
matrix of potentials. Thereafter with 
the interaction held fixed, we predict the spectra of the other 
mass-7 isobars, of which ${}^7$He and ${}^7$B are particle  unstable.
The MCAS method~\cite{Am03} appears particularly suited to  
this problem in that it describes the bound and resonant spectra of even-odd 
light nuclei in terms of nucleon-nucleus dynamics in the context of  
coupled-channel interactions.  Even a simplistic collective model, with 
geometric-type deformation for the structure of the nucleus, can give a very 
reasonable description of the compound system~\cite{Pi05}. 
That is so only if the non-local effects of Pauli blocking 
are taken into account in the nucleon-nucleus model Hamiltonian. 
With MCAS, using a collective model prescription for the input matrix of
interaction potentials, the Pauli exclusion principle can be taken into 
account by means of the Orthogonalizing Pseudo-Potential (OPP) 
method~\cite{Ca05}.  Single-particle and collective-model aspects then can be 
handled in an single approach for nuclear structure and scattering since
both sub-threshold and resonance states of the compound nucleus
can be determined.

Originally, the MCAS method was developed and tested in studies with
stable  light nuclei~\cite{Am03,Pi05,Sv06}. Subsequently the method was
developed  further to infer properties of a light nucleus that is outside
of the proton  drip line~\cite{Ca06}. Crucial to the outcome was the
introduction of a new  concept which combines collectivity and
single-particle dynamics, namely  {\em Pauli hindrance}.  This accounts
for the fact that the transition from Pauli-allowed to Pauli-blocked
single-particle orbits in shell-model structures is not a binary jump 
but might change gradually with mass.
The description in  terms of the OPP entails the non-local effects of
nucleon state  antisymmetrization, and appears perfectly suited 
parametrically to describe a smooth transitional situation~\cite{Ca06}.

In the following section, features of the MCAS method and of the
properties assumed for the target nuclei are discussed. 
This section also includes use of the MCAS method to analyze the spectra 
of ${}^7$Li and of ${}^7$Be as dicluster systems
of $\alpha$+${}^3$H and $\alpha$+${}^3$He respectively. 
A dicluster-model potential is generated that describes bound and 
resonance states.
The elastic
scattering cross sections of the cluster pair for energies to 14 MeV
are presented and discussed in that section as well. 
Thereafter, in 
Sec.~\ref{Results}, the results of our calculations of the coupled-channel
problems of a nucleon with a mass-6 nucleus are given and compared
with the known spectra of the mass-7 isobars. Finally, in Sec.~\ref{Conclude},
we give conclusions that may be drawn.

\section{The MCAS theory and some structure details}
\label{Theory}

As all details of the MCAS theory have been presented previously~\cite{Am03},
only salient features are reviewed herein.

\subsection{Application of MCAS to a nucleon-nucleus coupled-channel system}

We use a simple collective-model prescription for the matrix of interaction potentials
between a proton and $^6$He, similar to that used 
previously~\cite{Am03,Ca05,Pi05}, taking just a quadrupole deformation
(deformation parameter $\beta_2$) with a mix of central (0), spin-orbit ($ls$),
$l^2$ ($ll$), and spin-spin ($Is$) potential terms, and a Coulomb potential
from a uniformly charged sphere of radius $R_{c}$, {\it viz.}
\begin{eqnarray}
V_{cc'}(r) &=& V_0 v^{(0)}_{cc'}(r,\beta_2)
+ V_{ls} v^{(ls)}_{cc'}(r,\beta_2) 
\nonumber\\
&&\hspace*{1.0cm}  +
V_{ll} v^{(ll)}_{cc'}(r,\beta_2) + V_{Is} v^{(Is)}_{cc'}(r,\beta_2)
+ \delta_{cc'} V_{coul}(r, R_{c})\ .
\label{Vmat}
\end{eqnarray}
The channel index identifies the coupling of specific nucleon partial waves
to specific target states leading to each considered, and conserved, total 
spin-parity $J^\pi$ of the compound system, {\it viz.} $c \equiv \left[\left(l
\frac{1}{2}\right) j I: J\right]$, with parity $(-1)^l$ since we consider
only positive-parity target states in these applications.
It is important to note that the channel indices $c$ incorporate all relevant quantum 
numbers for a given single-particle state.

The model takes into account effects of core excitation and polarization by
allowing transitions from the ground state to the lowest two excited 
states which are assumed to have collective nature.  Therefore, to define the channel
space, we assume that there are three important states to consider in the
spectrum of  $^6$He.  They are the $0^+$
ground, a $2^+_1$ (1.78 MeV) first excited, and a $2^+_2$  (5.6 MeV) second
exited states. The ground and first excited states of ${}^6$He have been given those
spin  assignments~\cite{Ti02}, while the third is that expected by
a shell-model calculation~\cite{Ka00}.  Further, for simplicity, we
consider transitions between them to be effected  by one and the same
quadrupole operator though, in expansions, we take the
quadrupole deformation  to second order~\cite{Am03}. The basic
functional form of the channel-coupling  interactions is of Woods-Saxon
type.    

Starting from this local form in coordinate space, the full nuclear 
potential $\mathcal{V}$ contains, in addition, the highly nonlocal OPP term.
The final form is
\begin{equation}
\mathcal{V}_{cc'}(r,r')  =  V_{cc'}(r)\delta(r-r')  +  \lambda_c
    A_c(r) A_c(r')\delta_{cc'} .
\label{OPPeq}
\end{equation}
The function $A_c(r)$ represents the normalized radial part of the 
single-particle bound state in channel $c$, spanning the phase-space excluded 
by the Pauli principle.  The OPP method for treating the effects
of the Pauli-blocked states holds in the limit $\lambda_c \to \infty$, but it 
suffices to set $\lambda_c = 1000$~MeV to get a stable spectrum
where all forbidden states have been removed.  For Pauli-allowed states 
$\lambda_c = 0$. Pauli-hindered states are assumed when specific strengths of 
$\lambda_c$ are selected, greater than zero but much lower than $1$ GeV and 
typically a few MeV. Those strengths ($\lambda_c$) presently are treated as
free parameters.

\subsection{Aspects of structure}
If we consider $^6$He from the point of view of the simplest vibrational model,  
the two excited states are assumed, respectively, to be a one-phonon and a 
two-phonon excitation (both with L=2) from the ground.  As a consequence, and 
allowing for the scale factor (of 2) that differentiates basic probability 
expressions for one-phonon couplings between the states, this simple model 
predicts 
\begin{eqnarray*}
B(E2;2^+_2 \rightarrow 2^+_1)\  &=& 2\  
B(E2;2^+_1 \rightarrow {\rm g.s.)}\\
B(E2;2^+_2 \rightarrow {\rm g.s.)} &=& 0\ .
\end{eqnarray*} 
Neither are close to most observed results as, empirically,
\begin{equation}
0.5 \le \ \  {\cal R} =  \frac{B(E2;2^+_2 \rightarrow 2^+_1)} 
{B(E2;2^+_1 \rightarrow {\rm g.s.)}} \ \ \le 1.6\ .
\end{equation}

Wave functions for $^6$He have been  obtained from a complete 
$(0+2+4)\hbar\omega$ shell-model calculation~\cite{Ka00} in which the $G$-matrix 
interactions of Zheng~{\it et al.}~\cite{Zh95} were used. This
no-core model gave a spectrum with three low-lying states coinciding
with the known $0^+$ ground and the two excited (resonant) states having
centroids at  1.797 and 5.6 MeV. Of those the 1.797 MeV has been
assigned $2^+$ while the 5.6 MeV resonance is listed~\cite{Ti02} with
ambiguous spin-parities of $(0^+, 1^-, 2^+)$.  The shell model, as noted
previously, anticipates  $2^+_2$. All three states are  radioactive with
the 1.797 MeV state having a width of 113 keV. The width  of the 5.6
MeV state is uncertain. Other excited states are listed but lie much
higher in excitation~\cite{Ti02}, above 14 MeV.

Using bare charges and oscillator wave functions with an oscillator length of 
1.8 fm, the no-core shell model gave $B(E2)$ values for $\gamma$-decays in
${}^6$He  of 
\begin{eqnarray*}
B(E2;2^+_1 \rightarrow {\rm g.s.)} &=& 0.153\; \;{\rm e}^2 {\rm fm}^4\\
B(E2;2^+_2 \rightarrow 2^+_1)\  &=& 0.099\; \;{\rm e}^2 {\rm fm}^4\\
B(E2;2^+_2 \rightarrow {\rm g.s.)} &=& 0.036\; \;{\rm e}^2 {\rm fm}^4\ .
\end{eqnarray*}
Thus this shell-model calculation of $^6$He gives a ratio  ${\cal R} = 0.647$.
This lies near the lower limit of the empirical ratio range.
Also the shell model predicts that the 
$2_2^+$ state decays to the ground with a significant probability (23.5\%
of that of the $2_1^+$ state).
 
The isospin mirror, ${}^6$Be, is particle unstable (decaying to an
$\alpha$ and two  protons) and from the TUNL Data-Group
project~\cite{Ti02}  it is thought to have a resonant  $0^+$ ground
state and a first excited one, possibly $2^+$, centered 1.7 MeV above.
Nonetheless we take for ${}^6$Be the same shell-model spectroscopy of
${}^6$He,  under the assumption of  charge symmetry of the nuclear
force. Thus we assume a second $2^+$ excitation, centered around 
$~$5.6 MeV, also for the $^6$Be isotope. In the present analysis we treat
all nuclear (target) states, either ground or excited,  as stable.
In the
MCAS scheme unstable states could be accommodated in the formalism.
We plan to do so in the near future. Of course the Coulomb
interactions used in MCAS calculations take into account the change
from 2 to 4 protons and the OPP term refers to states forbidden in the
corresponding  mirror system.  The Coulomb radius was increased to 2.8 fm
in this case as well. 

As the shell-model calculations give for ${}^6$He (and of ${}^6$Be)
comparable admixtures  of pair re-coupling and pair breaking in the two 
$2^+$ states, in the collective model prescription of the input matrix
of interaction potentials, we have taken the two excitations to be equal 
mixtures of first
and  second order terms in the quadrupole-deformation parameter. 
The deformation $\beta_2$ is allowed to be a
variable parameter, to be adjusted along with the other parameters of the model 
interaction to best reproduce the known $^7$Li spectrum. 

\begin{table}
\begin{ruledtabular}
\caption
{\label{OMparams} Parameter values of the (negative parity) $p$-${}^6$He interaction.}
\begin{tabular}{ccccc}
Strengths & $V_0$ & $V_{\ell \ell}$ & $V_{\ell s}$ & $V_{Is}$ \\
\hline
 & $-$36.817 & $-$1.2346 & 14.9618 & 0.8511 \\
\hline
Geometry & $R_0$ fm; & $a$ fm;  & $R_c$ fm; & 
$\beta_2$\\
\hline
 & 2.8 & 0.88917 & 2.0 & 0.7298
\end{tabular}
\end{ruledtabular}
\end{table}

We have also  performed a shell-model calculation for $^7$Li using the 
$G$-matrix shell-model interaction of Zheng {\em et al.}~\cite{Zh95}. The
negative parity states were calculated within a complete
$(0+2+4)\hbar\omega$ model space, as was previously published~\cite{Ka97}, 
while the positive parity states were obtained using a
$(1+3+5)\hbar\omega$ model space. The only restriction in the latter
was the exclusion of the $5\hbar\omega$ 1p-1h components connecting
the $0p$ shell to the $0i$-$1g$-$2d$-$3s$ shell. 
The shell-model code OXBASH~\cite{Ox86} 
was used for all the calculations from which states
up to, and including $J = \frac{7}{2}$, were obtained. 
We will present and discuss later the results in comparison with those we obtain
using MCAS.

Of course, there are many other models
for the structure of these nuclei and of ${}^7$Li in particular; no-core shell 
models~\cite{No06} other than that we have used,  Green's function  Monte Carlo 
studies~\cite{Pi04}, 
and cluster model investigations~\cite{Wa85} to state a few.  We also stress the
complementary nature of these methods. Cluster and shell-model studies are
suitable, in general, for analyses of different data,  yet they can  also give
consistent descriptions of many nuclear properties, as has been pointed 
out recently~\cite{Za02}. 
That consistency is evident when one compares the electron scattering 
form factors for ${}^7$Li deduced from a cluster model~\cite{Wa85} and from
a no-core-shell-model calculation~\cite{Ka97}. Those models provide  equivalently
good fits to data to 2 fm$^{-1}$. 

Recently~\cite{Fo05},
a dicluster model of ${}^7$Li was used to study electromagnetic properties and break-up.
The $\alpha$-${}^3$H system is a single channel problem given the spectra of
the two nuclei involved.  
To reproduce the experimental energies of the four states in ${}^7$Li considered, 
the interaction strengths were adjusted in calculation of the $p$-wave and $f$-wave
functions of relative motion separately.  With those wave functions
a variety of data could be described ~\cite{Fo05}; most  of which  data
are sensitive primarily to the large radius properties of the wave functions.
In the following subsection, we develop a similar dicluster-model calculation
that reproduces the subset of mass-7 energy levels that significantly couples
to the cluster channel.

\subsection{Application to the mass-7 dicluster systems}

With the MCAS scheme, we have performed an equivalent dicluster calculation, identifying
the dicluster problem as a single-channel potential problem. 
We have ascertained a single potential that gives a set of compound states
in good agreement with some of those in the spectrum of ${}^7$Li~\cite{Ti02}.
The MCAS calculations were made without OPP. As a consequence,
there are two deeply bound spurious states
which have to be  neglected.  This is not an essential problem in single-channel
studies as, by construct, they are orthogonal to the other excited states.
When one considers a coupled-channel problem, however, that is no longer true and
due care of the Pauli principle is needed to ensure that all determined 
states have no spurious components~\cite{Ca05}.
Later we will incorporate a positive-parity interaction to analyze
scattering data. With that interaction, there are  also 
spurious sub-threshold positive-parity states in the evaluated  ${}^7$Li spectrum.
There are three such states of spin-parity $\frac{1}{2}^+$, $\frac{3}{2}^+$, 
and $\frac{5}{2}^+$, having energies of $-$11.1, $-$6.8, and $-$9.0 MeV relative 
to the $\alpha$+${}^3$H threshold for the interaction we have determined.  
These are indeed spurious as there is 
no known positive parity state in the spectrum below at least 11 MeV 
excitation~\cite{Ti02}. 

We first discuss results obtained using an  $\alpha$-${}^3$H potential acting
only in negative-parity states. 
A standard Woods-Saxon form~\cite{Am03}  was used with parameter values
\begin{equation}
\begin{array}{lll}
V_0 = -76.8\ {\rm MeV}\hspace*{1.0cm} & V_{ll} = 0.6\ {\rm MeV} &
\hspace*{1.0cm} V_{ls} = 1.7\ {\rm MeV}\\
R_0 = 2.39\ {\rm fm} & a = 0.68\ {\rm fm} &
\hspace*{1.0cm} R_c = 2.34\ {\rm fm}\\
\end{array}
\end{equation}
$R_c$ is the radius of a Coulomb sphere of charge.
As the $alpha$-particle is treated solely
as a $0^+$ state in this model, $V_{I\cdot s} = 0$.
For reference later, we define this (purely negative-parity) interaction
as Potential I.
This was the form that we found best reproduced the known energies
of the four physical states of interest in ${}^7$Li. Their values and 
resonance widths are given in  Table~\ref{alphat}.  
\begin{table}[h]
\begin{ruledtabular}
\caption{\label{alphat}
Spectral properties of ${}^7$Li  found using Potential I in the MCAS 
calculation of the $\alpha$+${}^3$H system.
The energies (center of mass, MeV) are relative to the 
$\alpha$-${}^3$H threshold.  The widths, given in brackets, are in keV.}
\begin{tabular}{ccccc}
$J^\pi$ & Exp. & Potential I \\
\hline
$\frac{3}{2}^-$ & spurious  & $-$29.4  \\
$\frac{1}{2}^-$ & spurious   & $-$27.8 \\ 
\hline 
$\frac{3}{2}^-$ & $-$2.47      & $-$2.47 \\
$\frac{1}{2}^-$ & $-$1.99      & $-$1.75  \\
$\frac{7}{2}^-$ & 2.18 (60)  & 2.12 (83)  \\
$\frac{5}{2}^-$ & 4.13 (918) & 4.17 (834)  \\
\end{tabular}
\end{ruledtabular}
\end{table}
It is important to note that, with the dicluster model, 
there are no other negative-parity states. 
But within the low-energy excitation 
range, a number of other states are known experimentally~\cite{Ti02}.

\subsubsection{$\alpha$ scattering from ${}^3$H}

Cross sections at select center of mass scattering angles
from the elastic scattering of $\alpha$-particles from a  
${}^3$H target for energies between 4 and 13.2 MeV have been measured
and a phase-shift analysis made~\cite{Sp67}. Three resonances were noted,
with the phase-shift analysis showing that the $\frac{7}{2}^-$ and 
first $\frac{5}{2}^-$ states were built from the relative $f$-wave
while the third, a weaker $\frac{5}{2}^-$, was
built from the relative $p$-wave in the scattering. In 
sequence, their centroids were put at lab. energies of 5.2,
9.8, and $\sim$ 11.5  MeV which link to the known state values in ${}^7$Li
at 4.65, 6.60 and 7.45 MeV excitation. The first two correspond to the
2.18 and 4.13 MeV states in respect to the $\alpha$-${}^3$H threshold.

\begin{figure}
\scalebox{0.5}{\includegraphics{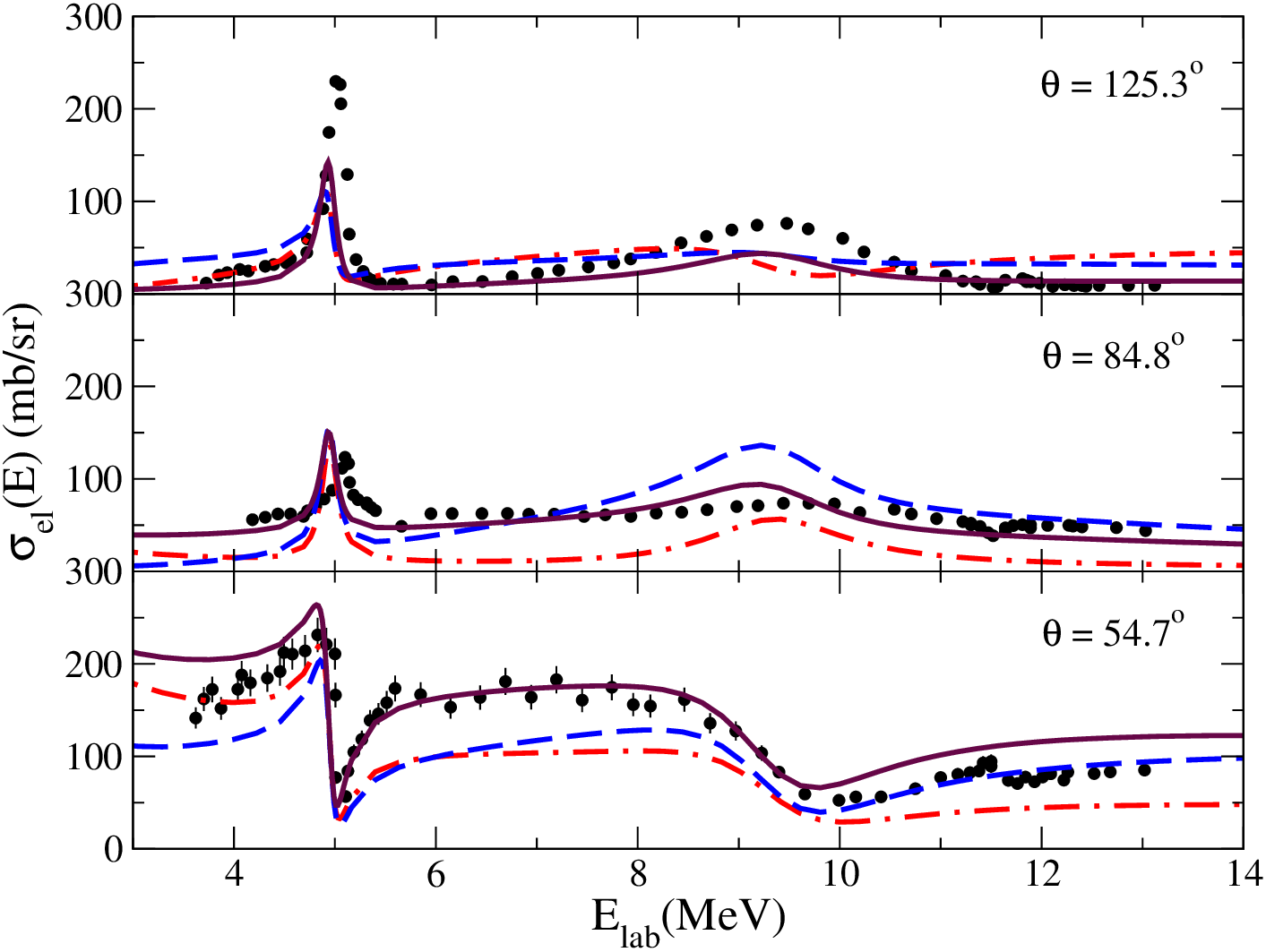}}
\caption{\label{fig1} (Color online)
Cross sections from ${}^3$H($\alpha,\alpha$)${}^3$H at the center of mass
scattering angles listed. Results differ by various amounts
of positive-parity contributions as described in the text.}
\end{figure}

As there are no known positive-parity states in the spectrum of ${}^7$Li,
one can only rely upon scattering data to assess the positive-parity 
$\alpha$-${}^3$H interaction. In Fig.~\ref{fig1}, a set of
results are given for three designated scattering angles at which data
have been obtained~\cite{Sp67}.
Results found using the Potential I interaction (no positive-parity
interaction) are displayed  by the dot-dashed curves.
The dashed curves depict results found  with Potential I and
assuming that the negative- and  positive-parity parameters are the same.
The solid curves result when that positive-parity central strength
is reduced to $-$70 MeV, which interaction we identify hereafter as
Potential II.

At the scattering angle of 54.7$^\circ$ 
the cross section found using Potential II 
reproduces  the data well at least to 10 MeV.
It is interesting to observe that 
a decrease in strength of the positive-parity interaction
actually enhances the cross-section results to achieve a
good comparison with data. 
This is a signature of strong interference effects between
partial-wave contributions. 
Preference for Potential II is also confirmed by the results
at the other scattering angles.

\subsubsection{The case of $\alpha$ + ${}^3$He and ${}^7$Be}

Assuming charge symmetry and changing only
the Coulomb-force details, we have used Potential I to describe 
the states of ${}^7$Be.
\begin{table}
\begin{ruledtabular}
\caption{\label{Be7-states}
Spectral properties found using the model Potential I in the 
MCAS method
for the  ${}^7$Be spectra from the  $\alpha$+${}^3$He system.
The energies (center of mass, MeV) are relative to the
$\alpha$-${}^3$He threshold.  The widths are in keV.}
\begin{tabular}{ccc}
$J^\pi$ & Exp. & Potential I \\
\hline
$\frac{3}{2}^-$ & spurious  & $-$28.0 \\
$\frac{1}{2}^-$ & spurious  & $-$26.4 \\
\hline
$\frac{3}{2}^-$ & $-$1.59 & $-$1.53\\
$\frac{1}{2}^-$ & $-$1.16 & $-$0.84 \\
$\frac{7}{2}^-$ & 2.98 (175) &  3.07 (180) \\
$\frac{5}{2}^-$ & 5.14 (1200) & 5.09 (1194) \\
\end{tabular}
\end{ruledtabular}
\end{table}
We have also changed
the Coulomb radius to 2.39 fm. That change has minimal
impact. Comparison with known state 
energies~\cite{Ti02} is given in Table~\ref{Be7-states}.
\begin{figure}[h]
\scalebox{0.5}{\includegraphics{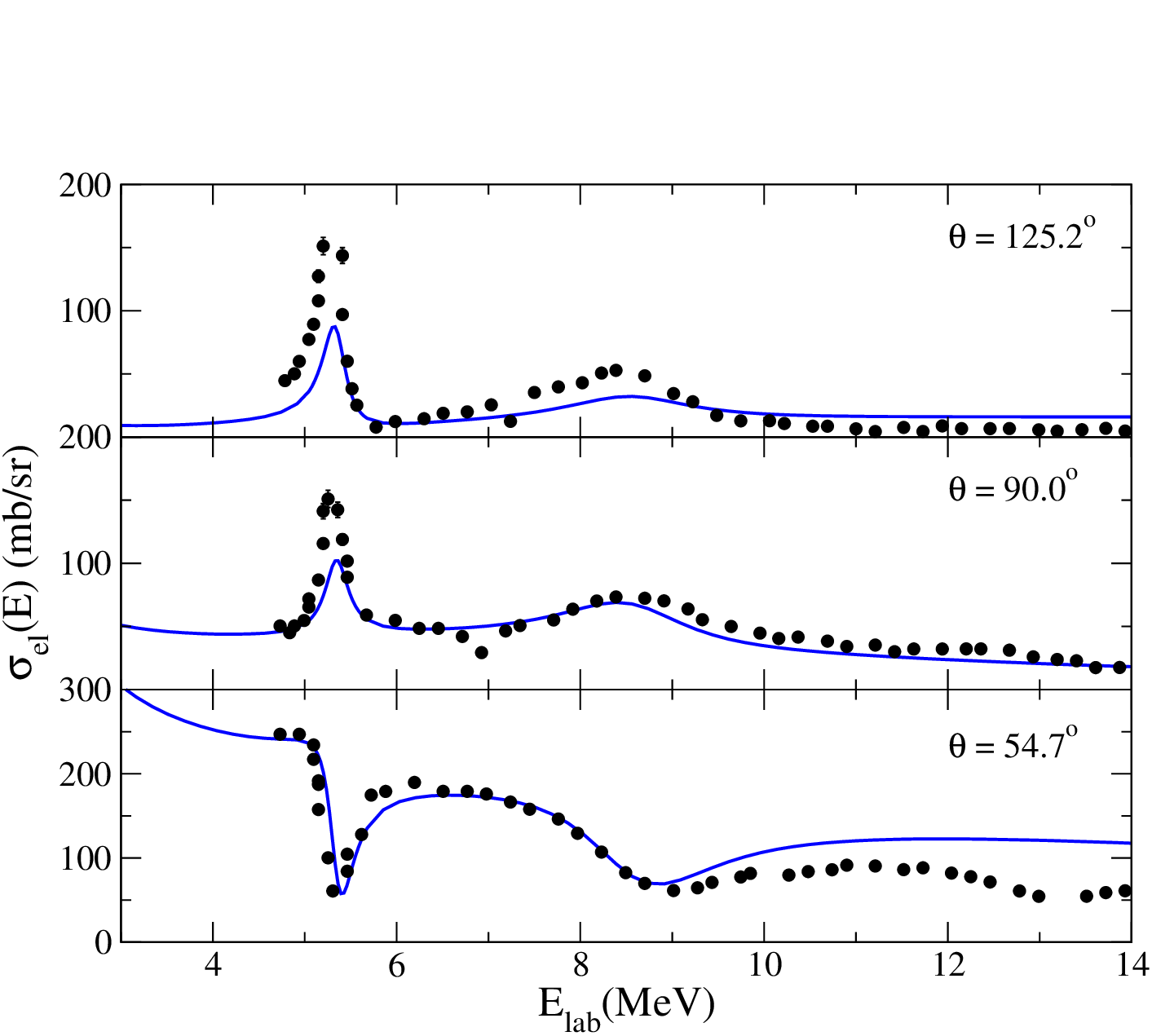}}
\caption{\label{fig2} (Color online)
Cross sections from $\alpha({}^3$He,${}^3$He$)\alpha$ at the center of mass
scattering angles listed. The curves are predictions made using the 
Potential II as the inter-nuclear interaction.}
\end{figure}
The energies of three of the four low lying states of ${}^7$Be are found within 100 keV 
of the experimental values and the fourth, the $\frac{1}{2}^-$ state, within 320 keV.
Even better are the widths for the two resonance levels being found to within 6 keV.
Then we used the MCAS results with Potential II  to form elastic scattering
cross sections at  three center of mass scattering angles 
($54.7^\circ, 90.0^\circ, 125.2^\circ$) at which data has been taken~\cite{Sp67}.
The results of those calculations are shown by the solid curves in Fig.~\ref{fig2}.
The comparisons with data are very good, adding confirmation to our definition
of the basic dicluster-interaction potential.


\section{results of nucleon+mass-6 nuclei calculations}
\label{Results}

Now we return to the description of $A=7$ nuclei in terms of a nucleon
plus mass-6 systems.
The ground state energies of the four nuclei of interest,
${}^7$He, ${}^7$Li, ${}^7$Be, and ${}^7$B lie
at 0.445, $-$9.975, $-$10.676, and  2.21 MeV with respect to the 
relevant nucleon-emission thresholds.

The known spectrum of ${}^7$Li~\cite{Ti02} consists exclusively of
negative-parity states so that provides no
information on the even-parity interactions. 
In fact, all mass-7 isobars have only negative-parity 
entries in their measured spectra up to the excitation energy studied. 
Since the MCAS method also can define scattering, 
it is hoped that low--energy ${}^6$He ion scattering from hydrogen may soon be
measured experimentally, not only to
ascertain resonances but also as the average
scattering cross section will be sensitive to
the positive-parity interaction in the $p$+${}^6$He system,
just as we have found using our dicluster-model potential. 
It is due to this lack of experimental information 
that we  restrict consideration hereafter to just the effects of
the odd-parity interaction.  

That all the states are of negative parity may be understood in the 
simplest shell picture  as being dominantly a capture of a $0p$ shell
nucleon upon the target states. To form a positive-parity state one has 
to have capture in the $1s$-$0d$ shells which requires 1$\hbar \omega$ additional 
energy. For these nuclei, that would be 10 MeV or greater.
The $0s$ shell is taken to be fully occupied and so no capture can be made into it.
Thus, in the MCAS approach for the mass-7 isobars, the $0s$ shell is to be explicitly 
Pauli-blocked. This is achieved by the OPP containing a term that prevents further 
proton- and neutron-occupancies in the  $0s$ configurations of the cores.

All results that we present here have been produced with the
potential parameter set given in Table~\ref{OMparams}. The components
of the potentials are identified in the first line of this table, and their
strengths, in MeV, are given in the second. The third line identifies the
geometry parameters whose values are listed in the last line.
Of the eight parameters of the table, the effective nucleon-nucleus 
radius $R_0$ and the corresponding charge radius $R_c$ has been set
consistently with shell-model information of $^6$He, corrected  due to
the size of the (impinging) proton. The remaining 6 parameters have been varied as
fit parameters to reproduce 6 out of the 11  known states of $^7$Li, 8 of which
are stable with respect to proton emission.
\begin{table}
\begin{ruledtabular}
\caption{\label{table-BOUND}
Experimental data and theoretical results for $^7$Li and $^7$Be states
(Energies are in MeV, widths are in keV). All energies are defined with thresholds,
$p$+${}^6$He = 9.975 MeV with respect to $^7$Li ground state  and
$n$+${}^6$Be = 10.676 MeV with respect to $^7$Be ground state.}
\begin{tabular}{c|cc|cc}
$J^\pi$  & \multicolumn{2}{c}{${}^7$Li} & \multicolumn{2}{c}{${}^7$Be} \\
\hline                                                  
 & Exp. & Theory & Exp. & Theory\\
\hline
$\frac{3}{2}^-$   & $-$9.975         & $-$9.975      &    $-$10.676          & $-$11.046       \\
$\frac{1}{2}^-$   & $-$9.497         & $-$9.497      &    $-$10.246          & $-$10.680       \\
$\frac{7}{2}^-$   & $-$5.323 [69] & $-$5.323      &    $-$6.106 [175]   & $-$6.409        \\
$\frac{5}{2}^-$   & $-$3.371 [918] & $-$3.371      &    $-$3.946 [1200]     & $-$4.497        \\
$\frac{5}{2}^-$   & $-$2.251 [80]  & $-$0.321      &    $-$3.466 [400]     & $-$1.597        \\
$\frac{3}{2}^-$   & $-$1.225 [4712] & $-$2.244      &          $--$         &    $--$         \\
$\frac{1}{2}^-$   & $-$0.885 [2752] & $-$0.885      &                       & $-$2.116        \\
$\frac{7}{2}^-$   & $-$0.405 [437] & $-$0.405      &    $-$1.406 [?]       & $-$1.704        \\
$\frac{3}{2}^-$   &      $--$        &    $--$       &    $-$0.776 [1800]     & $-$3.346        \\
$\frac{3}{2}^-$   &    1.265 (260)  & 0.704 (56) &     0.334 (320)       & $-$0.539        \\
$\frac{1}{2}^-$   &                     & 1.796 (1570) &                     & 0.727 (699)   \\
$\frac{3}{2}^-$   &  3.7 (800) ?$^a$     & 2.981 (990)&                 & 1.995 (231)   \\
$\frac{5}{2}^-$   &  4.7 (700) ?$^a$    & 3.046 (750)&                  & 2.009 (203)   \\
$\frac{5}{2}^-$   &                  & 5.964 (230)  &                       & 4.904 (150)   \\
$\frac{7}{2}^-$   &                  & 6.76  (2240)  &      6.5 (6500) ?$^b$       & 5.78  (1650)     \\
\end{tabular}					                          
$^a$ For these states spin and parity are unknown~\cite{Ti02}.\\
$^b$ Spin-parity of this state has been assigned as $\frac{1}{2}^-$~\cite{Ti02}.\\
\end{ruledtabular}
\end{table}

The fit procedure gave a  deformation parameter $\beta_2 = 0.7298$. This
is a large deformation, more than twice that inferred by using the
shell-model $B(E2)$ to determine a collective model value. The
diffusivity is larger than typically found with scattering from stable nuclei,
but that may simply reflect the neutron halo character of ${}^6$He.  The spectrum  
found from the MCAS
evaluations of  the $p$+${}^6$He system is compared 
in Table~\ref{table-BOUND} with  the
experimentally known  spectrum of ${}^7$Li~\cite{Ti02}.
Therein we also give the results  obtained for the mirror
case, ${}^7$Be,  which has been treated in the MCAS  formalism as an
$n$+${}^6$Be system. The energies are in MeV but the widths, shown in
brackets, are in keV.
The numbers in square brackets 
are the corresponding  experimental widths with respect to the $t$+$\alpha$
channel for $^7$Li, and $^3$He +$\alpha$ channel  for $^7$Be. Also the
decay into the channels $n$-$^6$Li and $p$-$^6$Li are included,
respectively, when these channels are open. Above the relevant zero-energy 
thresholds
the experimental widths are indicated in  round
brackets and are compared with values calculated
with MCAS. As the theoretical widths refer only to the
specific nucleon decay channels, they differ somewhat in significance with
respect to the experimental ones which include other break-up contributions.
Our evaluation produced 12 states to 15 MeV excitation in 
${}^{7}$Li. The lowest 9 match states match known spin-parity states in the 
spectrum. while the next three calculated levels are in the
energy region  in which two resonant states of undetermined spin-parity
are known.
 The matched states agree quite well in energies save for a crossing 
of the $\frac{3}{2}^-|_2$ with the $\frac{5}{2}^-|_2$ states. 
A measure of the quality of result is that the  
mean square error, calculated over the 8 states in ${}^7$Li  below the 
proton emission threshold, is
\begin{equation}
\mu= \frac{1}{N} \sqrt{\sum (E_{th}-E_{exp})^2} = 0.2728 \ {\rm MeV}.
\end{equation}
By including the resonances, assuming the assignments, the mean square error 
remains good, namely $\mu = 0.2966$ MeV.
Also the agreement between widths for the resonances is 
satisfactory given that no adjustments have been made to get a
better fit to them. But the experimental widths have various components.
The ground state in ${}^7$Li is stable and the first excited decays 
only electromagnetically. The next two can decay also by emission 
of a triton or an $\alpha$ particle as the threshold for that decay
is 2.467 MeV above the ${}^7$Li ground state value. The next states in 
the spectrum can also decay by neutron emission as the $n$+${}^6$Li threshold 
lies 7.25 MeV above ${}^7$Li ground.

\begin{figure}
\scalebox{0.5}{\includegraphics{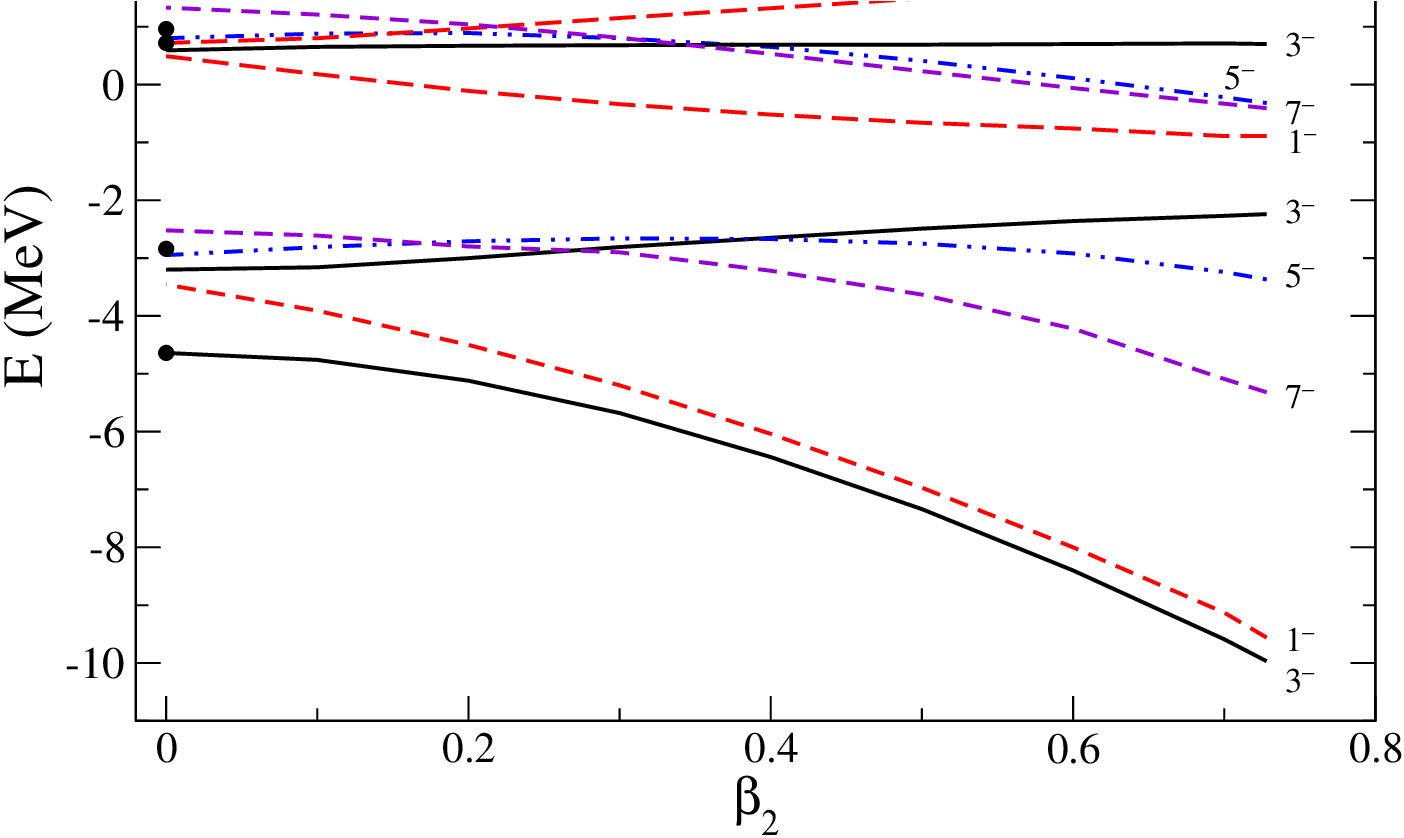}}
\caption{\label{fig3} (Color online)
The calculated energy spectrum of $^7$Li as $\beta_2$ varies.}
\end{figure}
To interpret the structure of the evaluated spectra,
we follow the procedure used previously~\cite{Pi05,Ca05} of allowing the deformation to
decrease to zero. The variation of the spectrum is shown in Fig.~\ref{fig3}.
Each state is identified on the right by twice its spin and its parity.
The dots show the degenerate spectrum values when $\beta_2 = V_{Is} = 0$.
States cluster as the deformation vanishes with the residual separation
of groups due to the effect of the (target)spin-(proton)spin interaction
$V_{Is}$ being finite. Evidently, the individual states retain an element 
of their zero deformation grouping for all of them track smoothly with $\beta_2$
and do not cross any member of another group.
Some states within each group do cross (interchange
their order in the spectrum as $\beta_2$ decreases) when $\beta_2$ lies in the 
range 0.2 to 0.35. Nevertheless, the mixing of basis states is evident from
the rapid divergence of the members of the two primary groups with 
increasing $\beta_2$.

\begin{figure}
\scalebox{0.6}{\includegraphics{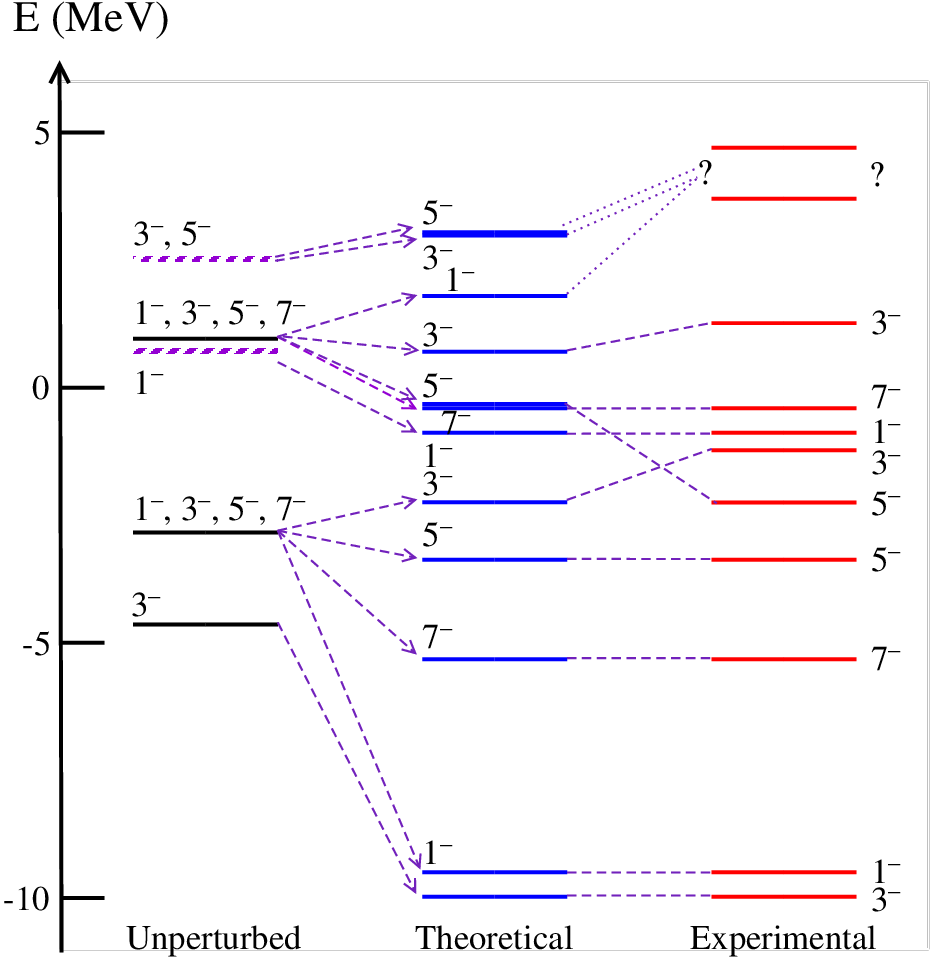}}
\caption{\label{fig4} (Color online)
The calculated energy spectrum of $^7$Li compared to the experimentally known 
one.  The left column is the zero deformation and  $V_{Is} = 0$ result. Spin-parities 
to the top most experimental levels have not been assigned}
\end{figure}
Setting that interaction strength $V_{Is} = 0$ at zero deformation gives 
the spectrum of degenerate states shown in the leftmost column of 
Fig.~\ref{fig4}. 
The states are identified by their value of (2$\times$spin)-parity.
The mapping of our results against the experimental spectrum discussed
above is clear on comparing the central (theoretical) against the
empirical (rightmost) spectra.  Comparing now with the zero deformation
results, it is clear that the dominant term in the ground state specification
is the coupling of a $0p_{\frac{3}{2}}$ proton  to the ground state 
of the target. Of course with deformation non-zero, the actual description
is a linear combination of all such basis $\frac{3}{2}^-$ states.
The next four states in the spectrum dominantly are formed by coupling
a $0p_{\frac{3}{2}}$ proton  to the first excited $2^+$ state of the target 
as the energy gap in the unperturbed spectrum is 1.78 MeV. The next state
with spin-parity $\frac{1}{2}^-$ is a member of the quartet of states
built primarily  from the coupling of a  
$0p_{\frac{3}{2}}$ proton  to the second  excited $2^+$ state of the target.
Then there is another $\frac{1}{2}^-$ state which resolves to the unique
coupling of a $0p_{\frac{1}{2}}$ proton with the ground state when
the deformation vanishes. Its energy (0.73 MeV) in that limit lies
$5.36$ MeV above the ground state value and that is the spin-orbit 
splitting of the single proton states in this model.
Finally  there is the doublet of states formed by coupling 
a $0p_{\frac{1}{2}}$ proton  to the first excited $2^+$ state of the target.
Confirmation of these assignments comes from tracking the widths of 
resonances found as $\beta_2$ decreases. For those states that are in, or move  
into, the continuum with decreasing $\beta_2$, their centroids and widths 
are given in Table~\ref{Li7-widths}. In this, the subscript $n$ designates
the rank of the state in the theoretical spectrum, while the letter `m' in the
brackets, which otherwise contains the width, signifies a value less than
1 eV. Usually it is much less than that.
\begin{table}[ht]
\begin{ruledtabular}
\caption{\label{Li7-widths}
Centroids (MeV) and widths (keV) of resonance states in ${}^7$Li as $\beta_2$ decreases. }
 \begin{tabular}{lllllllll}
 & \multicolumn{8}{c}{$\beta_2$ for ${}^7$Li} \\
\hline                                                  
$\left.J^\pi\right|_n$ & 0.7 & 0.6 & 0.5 & 0.4 & 0.3 & 0.2 & 0.1 & 0.0 \\
\hline
$\left.\frac{1}{2}\right|_2$ & & & & & & & 0.18 (1) & 0.49 (m)  \\
$\left.\frac{7}{2}\right|_2$ & & & 0.23 (m) & 0.53 (m) & 0.81 (m) & 1.04 (m) &
 1.21 (m) & 1.33 (m) \\
$\left.\frac{5}{2}\right|_2$ & & 0.11 (m) & 0.41 (m) & 0.65 (m) & 0.82 (m) & 0.89 (m) & 0.80 (m) & 0.80 (m)\\
$\left. \frac{3}{2}\right|_3$ & 0.71 (54) & 0.70 (42) & 0.69 (30) & 0.69 (18) & 0.68 (10) 
& 0.67 (4) & 0.65 (m) & 0.59 (m) \\
$\left.\frac{1}{2}\right|_3$ & 1.76 (1510) & 1.63 (1303) & 1.49 (1180) & 1.32 (858) & 1.15 (652) &
 0.97 (652) & 0.80 (334) & 0.72 (418)\\
$\left.\frac{3}{2}\right|_4$ & 2.96 (900) & 2.90 (848) & 2.83 (780) & 2.76 (680) & 2.69 (652) & 
2.65 (600) & 2.63 (580) & 2.66 (610)\\
$\left.\frac{5}{2}\right|_3$ & 3.04 (740) & 2.99 (700) & 2.92 (646) & 2.84 (584) & 2.75 (526) & 
2.66 (480) & 2.59 (440) & 2.53 (398) \\
\end{tabular}
\end{ruledtabular}
\end{table}
The quartet of states,  all having widths less than 1 eV, coincide with the zero deformation
and $V_{Is}= 0$ degenerate set having an energy of 0.96 MeV. They are built upon  a single particle
bound in the continuum. The $\frac{1}{2}^-$ state that tends to 0.724 MeV, and the doublet 
that tends 
to a degenerate value of 2.52 MeV, are built upon single particle resonances  in the continuum
whose inherent width is 580 keV.

The same nucleon-nucleus matrix of interactions, but with no Coulomb terms,
was used to evaluate the spectrum of the isospin mirror system $n$+${}^6$Be.
Again in making comparison with  the experimental spectra for ${}^7$Be,
one must bear in mind the threshold energies of reactions, ${}^3$He
+ $\alpha$ of 1.586 MeV, $p$+${}^6$Li of 5.806 MeV, and $n$+${}^6$Be at 
10.676 MeV. Our predicted spectrum of ${}^7$Be is compared with the 
known values also in Table~\ref{table-BOUND} and again
the round brackets around the calculated widths are solely for neutron emission.
The five lowest lying states in the known spectrum compare reasonably
with the MCAS values. The calculations give more states than are known to
date above an excitation energy of $\sim$ 8.5 MeV in ${}^7$Be, and there are 
a few crossings. But the result is a limited one in that it is predicated upon 
charge symmetry and the simple  collective model prescription. Studies using other
model prescriptions are in progress.
\begin{table}
\begin{ruledtabular}
\caption{\label{table-UNBOUND}
Experimental data and theoretical results for $^7$He and $^7$B states.
All energies are in MeV and relate to thresholds of $-$0.445 MeV for $n$+${}^6$He
and of $-$2.21 MeV for $p$+${}^6$Be.}
\begin{tabular}{c|cc|cc}
$J^\pi$  & \multicolumn{2}{c}{${}^7$He} & \multicolumn{2}{c}{${}^7$B} \\
\hline                                                  
 & Exp. & Theory & Exp. & Theory\\
\hline
$\frac{3}{2}^-$   & 0.445 (150)     & 0.43 (100)  &    2.21 (1400)    & 2.10 (190)       \\
$\frac{7}{2}^-$   &  $--$            & 1.70 (30) &                & 3.01 (110)     \\
$\frac{1}{2}^-$   &  1.0 (750) ?$^a$ & 2.79 (4100)  &             & 5.40 (7200)      \\
$\frac{5}{2}^-$   & 3.35 (1990)     & 3.55 (200)  &                & 5.35 (340)      \\
$\frac{3}{2}^-$   & 6.24 (4000) ?$^b$  & 6.24 (1900)  &                     &                  \\
\end{tabular}					                          
$^a$ Observed very recently and interpreted as a $\frac{1}{2}^-$ state.  \\
$^b$ Spin-parity of this state is unknown\\
\end{ruledtabular}
\end{table}
 
The spectrum of ${}^7$He, so far as it is known experimentally, 
has three resonant states with only the ground being quite narrow.
There is a claim~\cite{Me02} of a fourth resonance $\frac{1}{2}^-$ at $\simeq$ 
1 MeV above threshold which we include in Table VI.
Our calculation puts it higher in energy,
as do shell-model calculations.
Additionally we find a narrow $\frac{7}{2}^-$ resonance at 1.7 MeV excitation
that has not been observed.  
In the MCAS calculations of the $^7$He nucleus,
one must introduce a Pauli-hindrance effect
on the $0p$ shells. This effect produces a ground state that is unbound with 
respect to neutron emission. Specifically,
one must invoke an hindrance of both the 
$0p_\frac{3}{2}$ and $0p_\frac{1}{2}$ shells.
With this system, such effects might be a reflection of an  exotic and 
non-compact structure  ($^6$He)
being used as a basis in the channel coupling. 
A similar discussion applies also for a proton coupled to $^6$Be  states.
However, only the ground state of $^7$B is known and it is encouraging
that the MCAS calculation has found that resonance
energy accurate to 5\%. As shown in Tab. VI, we also predict three 
more resonances, two of which 
have widths sufficiently narrow to be detected in experiments.

With the same  nucleon-nucleus interaction, fixed by properties of ${}^7$Li,
and only modified by a Coulomb field, we describe spectra of two unbound nuclei 
$^7$He and $^7$B. 
The calculations of these differ from those for the isospin mirrors, 
${}^7$Li and ${}^7$Be, in
regard to  the OPP terms. While Pauli blocking of the $0s_{\frac{1}{2}}$ 
shell is
common to all four nuclides, the two particle-unstable ones
have additional OPP terms responsible for Pauli hindrance
in the $0p_\frac{3}{2}$ and $0p_\frac{1}{2}$ shells. The parameter
$\lambda_c$ was set as 17.8 MeV for the $0p_\frac{3}{2}$ shell
in each of the three target states considered,
while for the $0p_\frac{1}{2}$ shell, $\lambda_c$ was set as 
36.0 MeV for the $0^+$ g.s.,
but as 5.8 MeV for the two excited $2^+$ states.
The same hindrance effects were used for both $^7$He and $^7$B.

\begin{figure}
\scalebox{0.6}{\includegraphics{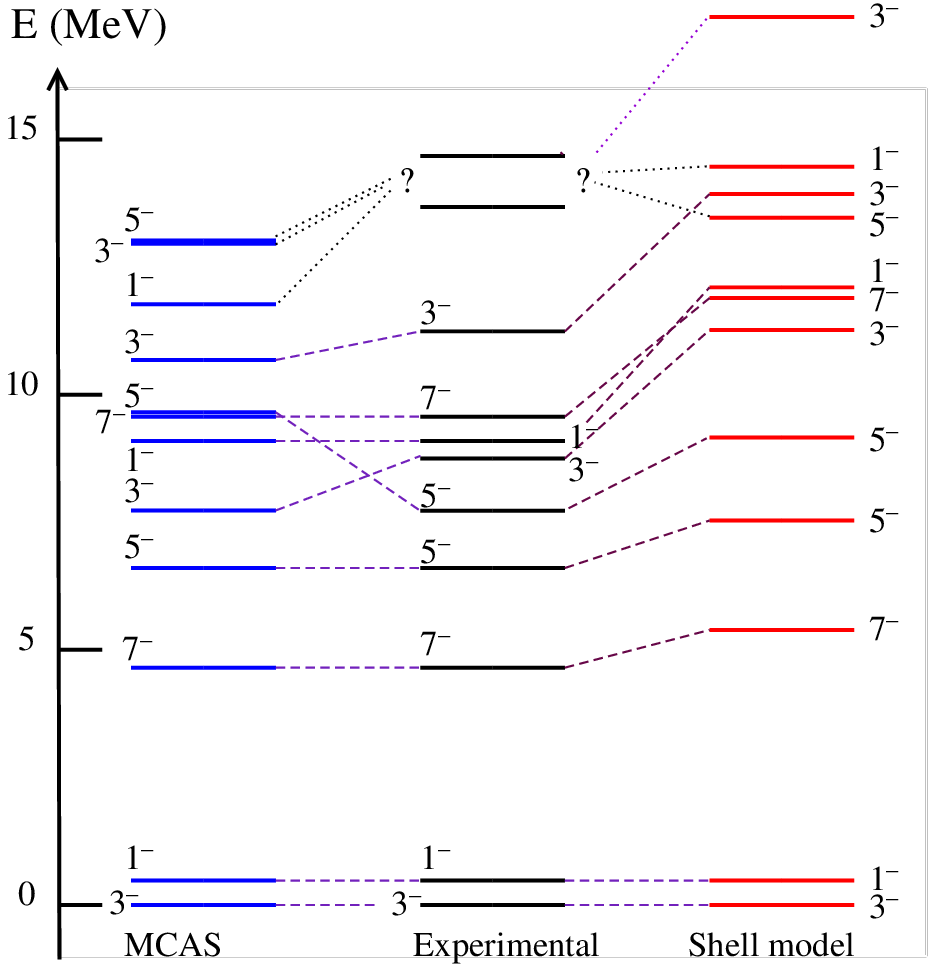}}
\caption{\label{fig5} (Color online)
The calculated energy spectra of $^7$Li compared to the experimentally known 
one.  The left and right columns are the MCAS and the no-core-shell-model results
respectively.  The known spectrum~\cite{Ti02} is given in the middle column.}
\end{figure}

Finally we compare the MCAS results for ${}^7$Li with both the
known spectrum and with that determined from our no-core-shell-model
calculations. Those spectra to 15 MeV excitation  are shown in Fig.~\ref{fig5}
and each state identified by twice its spin and its parity.
In both calculated spectra there is a high lying triplet of states, the 
$\frac{1}{2}^-|_3, \frac{3}{2}^-|_4,$ and $\frac{5}{2}^-|_3 $ that 
coincide with a doublet of resonances of unknown spin-parity in
the known spectrum~\cite{Ti02}.
All lower excitation states have defined spin-parity and are paired 
to ones in both  calculated spectra. The shell-model results  
reproduce the known spectrum~\cite{Ti02} quite well and that calculation
also 
found the lowest energy positive parity state to be a $\frac{1}{2}^+$ state
at 33.7~MeV excitation. 
The sequencing of the states in the spectrum given
by the shell-model calculation  is also very good with one minor cross
over of the $\frac{7}{2}^-|_2$ and $\frac{1}{2}^-|_2$ states, and a larger
cross over only occurring with the $\frac{5}{2}^-|_3$ at $\sim 15$ MeV excitation.
As noted there is a 1:1 correspondence between states in the MCAS and shell-model
spectra with the MCAS spectra (of states not used in determination of the
$V_{cc'}(r)$) being slightly compressed while the shell-model spectra  is
slightly expanded in energies in comparison to the known states. This, we
believe could be evidence of the need for more collectivity in the shell-model
description and a softening of that given by MCAS.

\section{Conclusions}
\label{Conclude}

We have used a collective-model prescription of a three-state 
($0^+$ ground, $2^+_1$, and $2^+_2$) spectrum for the isospin-mirror nuclei, 
${}^6$He and ${}^6$Be, in forming the coupling interactions with an extra nucleon
to yield the bound and resonant spectra of the mass-7 isobars. 
We used only a quadrupole deformation but chose parameters consistent
with the mass-6 targets having extended (halo) nucleon distributions.
We have also used a dicluster-model potential to assess states in ${}^7$Li and ${}^7$Be
that can have strong coupling in the cluster-cluster channels,
${}^3$H+$\alpha$ and ${}^3$He+$\alpha$.
The first four
levels of the nuclei are well reproduced as are the widths
of the $f$-wave resonances.  With this interaction,
the low-energy elastic scattering cross sections of the clusters
are also well reproduced.   However, the dicluster model gives no other state 
in the spectra, while a number of other states have been observed. 
In contrast, a complete reproduction
of the bound and resonance levels for all mass-7 isobars was found 
from the coupled-channel solutions of the nucleon-mass-6 systems when 
a single, fixed, nucleon-nucleus interaction was used. 
Specifically we found very good
reproductions of the spectra of the stable isobars, ${}^7$Li (from $p$+${}^6$He)
and ${}^7$Be (from $n$+${}^6$Be).  The other two isobars, ${}^7$He
(from $n$+${}^6$He) and ${}^7$B (from $p$+${}^6$Be) are, as known~\cite{Ti02},
particle unstable. 
The MCAS predictions for their (resonant) ground states
is consistent with the available data once Pauli-hindrance in the $0p$ shells 
is invoked.
Other yet to be discerned resonances are predicted, suggesting a
complex scenario of low-lying odd-parity resonances.

\begin{acknowledgments}
This research was supported by 
the Italian MIUR-PRIN Project      ``Struttura Nucleare e Reazioni Nucleari'' 
and by the Natural Sciences and
Engineering Research Council (NSERC), Canada.
K. A. and D. v.d. K. gratefully acknowledge the support and hospitality of the I.N.F.N.
(section Padova) and the University of Padova during visits in which
this research was developed, as do L. C. and J. P. S. for that given by the University 
of Melbourne during their visits.
\end{acknowledgments}

\bibliography{settebello11}

\end{document}